\documentclass[a4paper]{article}
\usepackage{ae}
\usepackage[english]{babel}
\usepackage{latexsym}
\usepackage{amssymb,amsmath}
\usepackage{epsf,graphicx}
\newtheorem{defi}{Definition}[section]
\newtheorem{prop}{Proposition}[section]
\newtheorem{thm}{Theorem}[section]

\newtheorem{lem}[thm]{Lemma}

\begin{document}

\title{\textbf{Good Illumination of Minimum Range}
\footnote{M. Abellanas, G.Hern\'andez and B. Palop supported by
grant TIC2003-08933-C02-01, CAM S0505/DPI/023 and MEC-HP2005-0137,
A. Leslie supported by CEOC through \emph{Programa} POCTI, FCT,
co-financed by EC fund FEDER and by Acção No. E-77/06 F. Hurtado
supported by Projects MCYT BFM2003-00368 and GenCat2005SGR00692,
I. Matos supported by CEOC through \emph{Programa} POCTI, FCT,
co-financed by EC fund FEDER and partially supported by Calouste
Gulbenkian Foundation and by Acção No. E-77/06}}

\date{}

\maketitle

\author{ M. Abellanas $^{\dag}$ $^{1}$, A. Bajuelos $^{2}$, G.
Hernández $^{1}$}
\author{F. Hurtado $^{3}$, I. Matos $^{2}$, B. Palop $^{4}$}

\author{ $^{1}$ \small{Universidad Politécnica de Madrid,}}
\author{ $^{2}$ \small{Universidade de Aveiro,}}

\author{ $^{3}$ \small{Universitat Politècnica de Catalunya,}}
\author{ $^{4}$ \small{Universidad de Valladolid}}

\author{ $^{\dag}$ \small{{\tt mabellanas@fi.upm.es}, corresponding
author}}

\begin{abstract}
A point $p$ is \mbox{1-well} illuminated by a set $F$ of $n$ point
lights if $p$ lies in the interior of the convex hull of $F$. This
concept corresponds to $\triangle$-guarding \cite{SE03} or
well-covering \cite{EHM04}. In this paper we consider the
illumination range of the light sources as a parameter to be
optimized. First, we solve the problem of minimizing the light
sources' illumination range to \mbox{1-well} illuminate a given
point $p$. We also compute a minimal set of light sources that
\mbox{1-well} illuminates $p$ with minimum illumination range.
Second, we solve the problem of minimizing the light sources'
illumination range to \mbox{1-well} illuminate all the points of a
line segment with an $\mathcal{O}(n^2)$ algorithm. Finally, we
give an $\mathcal{O}(n^2 \log n)$ algorithm for preprocessing the
data so that one can obtain the illumination range needed to
\mbox{1-well} illuminate a point of a line segment in
$\mathcal{O}(\log n)$ time. These results can be applied to solve
problems of \mbox{1-well} illuminating a trajectory by approaching
it to a polygonal path.
\end{abstract}


{\small
\section{Introduction and definitions}

Visibility or illumination has been the main topic for a lot of
different works but most of them cannot be applied to real life,
since they deal with ideal concepts. For instance, light sources
have some restrictions since they cannot illuminate an infinite
region as their light naturally fades as the distance grows. As
well as cameras or robot vision systems, both have severe
visibility range res\-tric\-ti\-ons because they cannot observe
with sufficient detail far away objects. We present some of these
illumination problems adding several restrictions to make them
more realistic, each light source has a limited illumination range
so their illuminated regions are de\-li\-mi\-ted. We use a limited
visibility definition due to Ntafos \cite{SN92} as well as a new
concept related to this type of problems, the \emph{t-good
illumination} due to Canales et. al \cite{ACH04,SC04}. This last
concept tests the light sources' distribution in the plane. If
they are somehow surrounding the object we want to illuminate,
there is a big chance it is $1$-well illuminated (\mbox{1-good}
illumination is also known as $\triangle$-guarding \cite{SE03} or
well-covering \cite{EHM04}).

This paper is solely focused in an optimization problem related to
limited \mbox{1-good} illumination. We propose the linear
algorithm \emph{MER-Point} to calculate the Minimum Embracing
Range (MER) of a point in the plane and it also solves the
decision problem. We move on to the computation of the MER of a
line segment. In order to do this, we propose an algorithm that
takes advantage of the Parametric Search \cite{NM79,NM83} and runs
in $\mathcal{O}(n^2)$ time. Our last algorithm is the main result
in this paper as it computes the E-Voronoi diagram \cite{ACHMP05}
restricted to a line segment which allows us to obtain the
illumination range needed to \mbox{1-well} illuminate a query
point of the line segment in $\mathcal{O}(\log n)$ time.

Let $F$ be a set of $n$ light sources in the plane that we call
sites. Each light source $f_i \in F$ has limited illumination
range $r > 0$, this is, they can only illuminate objects that are
within the circle centered at $f_i$ with radius $r$. As we only
consider \mbox{$1$-good} illumination, throughout this paper we
will refer to it just as illumination. The first two problems
minimize the light sources' range while illuminating certain
objects (points and line segments). On the third one, we also try
to answer efficiently which light source embraces any given point
in a line segment, this is, we want to compute the E-Voronoi
Diagram restricted to a line segment. The next definitions follow
from the notation introduced by Chiu and Molchanov \cite{CM03},
where CH$(F)$ denotes the convex hull of the set $F$.

\begin{defi}
A set of points $F$ is called an embracing set for a point $p$ in
the plane if $p$ lies in the interior of the $\mathrm{CH}(F)$.
\end{defi}

\begin{defi}
A site $f_i \in F$ is an embracing site for $p$ if $p$ is an
interior point of the convex hull formed by $f_i$ and by all the
sites of $F$ closer to $p$ than $f_i$.
\end{defi}

Following this definition, there may be more than one embracing
site for a point $p$. Since we are trying to minimize the light
sources' illumination range, in fact, we are trying to compute the
closest embracing site for point $p$ (see Figure \ref{EmbSites}).
The NNE-graph \cite{CM03} consists of a set of vertices $V$ where
each vertex $v \in V$ is connected to its first nearest neighbour,
its second nearest neighbour, ..., until $v$ is in the convex hull
of its nearest neighbours. Chan et al. \cite{MC04} present several
algorithms to construct the NNE-graph.

\begin{figure}[!h]
\centering
\includegraphics[scale=0.8]{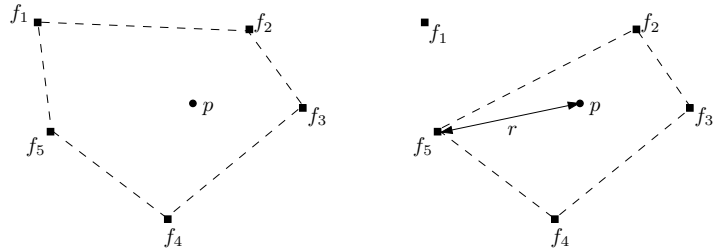}
\caption{The light sources $f_1$ and $f_5$ are em\-bra\-cing sites
for point $p$. The light source $f_5$ is the closest embracing
site for $p$ with illumination range $r=d(p,f_5)$.}
\label{EmbSites}
\end{figure}

\begin{defi}
Let $F$ be a set of $n$ light sources in the plane. We call
Clo\-sest Em\-bra\-cing Triangle for a point $p$,
$\mathrm{CET}(F,p)$, to a set of three light sources of $F$
containing $p$ in the interior of the triangle they define and
where one of the three light sources is the closest embracing site
for $p$.
\end{defi}

Since the set of light sources is always $F$, CET($F,p$) is
shortened to CET($p$) in this paper.

\begin{defi}[\cite{SC04}] Let $F$ be a set of $n$ light sources in
the plane. We say that a point $p$ in the plane is $t$-well
illuminated by $F$ if every open half-plane containing $p$ in its
interior, contains at least $t$ light sources of $F$
illu\-mi\-na\-ting $p$.
\end{defi}

This definition tests the light sources' distribution in the plane
so that the greater the number of light sources in every open
half-plane containing the point $p$, the better the illumination
of $p$. This concept can also be found under the name of
$\triangle$-guarding \cite{SE03} or well-covering \cite{EHM04}.
The motivation behind this definition is the fact that, in some
applications, it is not sufficient to have one point illuminated
but some neighbourhood of it \cite{EHM04}.

Let $C(f_i,r)$ be the circle centered at $f_i$ with radius $r$ and
let $A_r(f_1, f_2, f_3)$ denote illuminated area by the light
sources $f_1, f_2$ and $f_3$ (see Figure \ref{1GoodIl}(a)). It is
easy to see that $A_r(f_1, f_2, f_3)= C(f_1,r) \cap C(f_2,r) \cap
C(f_3,r)$. We use $A_r^E(f_1, f_2, f_3)=A_r(f_1, f_2, f_3) \cap
\mathrm{int}(\mathrm{CH}(f_1, f_2, f_3))$ to denote the
illuminated area embraced by the light sources $f_1, f_2$ and
$f_3$.

\begin{figure}[!h]
\centering
\includegraphics[scale=0.8]{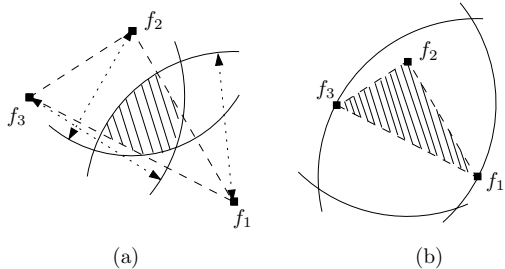}
\caption{(a) $A_r^E(f_1, f_2, f_3)$ is the shaded open area, so
e\-ve\-ry point that lies inside it is \mbox{1-well} illuminated
by the light sources $f_1, f_2$ and $f_3$. (b) All the interior
points of the $\mathrm{CH}(f_1, f_2, f_3)$ are \mbox{1-well}
illuminated since $r \geq \max \{d(f_i,f_j), f_i\neq f_j \in
\{f_1, f_2, f_3\} \}$.} \label{1GoodIl}
\end{figure}

\begin{defi}
We say that a point $p$ is \mbox{$1$-well} illuminated by the
light sources $f_1, f_2$ and $f_3$ if $p \in A_r^E(f_1, f_2, f_3)$
for some range $r>0$.
\end{defi}

\begin{defi}
Given a set $F$ of $n$ light sources, we call Minimum Embracing
Range to the minimum range needed to \mbox{1-well} illuminate a
point $p$ or a set of points $S$ in the plane, respectively
$\mathrm{MER}(F,p)$ or $\mathrm{MER}(F,S)$.
\end{defi}

Since the set $F$ is clear from the context, we will use ``MER of
a point $p$" instead of MER($F,p$) and ``MER of a set $S$" instead
of MER($F,S$). Once we have found the closest embracing site for a
point, its MER is given by the distance between the point and its
closest embracing site. So we know that a point is \mbox{1-well}
illuminated if all the light sources' illumination range is, at
leat, the same value as MER. In the next section we focus on our
solution to obtain the closest embracing site of $F$ for a point
$p$ as well as one CET($p$).

\section{Minimum Embracing Range of a Point}
\label{MERPoint}

Let $F$ be a set of $n$ light sources in the plane and $p$ a point
we want to \mbox{1-well} illuminate. The objective of this section
is to compute the value of the MER of $p$, $r_m$, as well as one
CET($p$). The closest embracing site for $p$ can be obtained in
linear time using the NNE-graph by Chan et. al \cite{MC04}. The
algorithm we present in this section also has the same time
complexity.

\textbf{The MER-Point Algorithm}

First of all, we compute the distances from $p$ to all the light
sources. Afterwards, we compute the median of all the distances in
linear time \cite{MB73}. Depending on this value, we can split the
light sources in two halves: the set $F_c$ that contains the
closest half to $p$ and the set $F_f$ that contains the furthest
half. We check whether $p \in \mathrm{int}(\mathrm{CH}(F_c))$,
what is equivalent to test if $F_c$ is an embracing set for $p$.
If the answer is negative, we recurse adding the closest half in
$F_f$. Otherwise, we recurse halving $F_c$. This logarithmic
search runs until we find the light source $f_p \in F$ and the
subset $F^E\subseteq F$ such that $p\in \mathrm{int}
(\mathrm{CH}(F^E))$ but $p\notin \mathrm{int} (\mathrm{CH}(F^E
\setminus \{f_p\})$. The light source $f_p$ is the closest
embracing site for $p$ and its MER is $r_m=d(f_p,p)$.

On each recursion, we have to check whether $p \in \mathrm{int}
(\mathrm{CH}(F')), F'\subseteq F$. This can be done in linear time
\cite{NM83-2} if we choose carefully the set of points so that
each point is studied only once. As soon as we have computed
$f_p$, we can find the two other vertices of a CET$(p)$ in linear
time as follows. Consider the circle centered at $p$ of radius
$r_m$ and the line $pf_p$ that splits the light sources inside the
circle in two sets. Note that if $f_p$ is the closest embracing
site for $p$ then there is a semicircle empty of other light
sources than $f_p$. A CET$(p)$ has $f_p$ and two other light
sources in the circle as vertices. Actually, any pair of light
sources $f_l, f_r$ such that each lies on a different side of the
line passing through $p$ and $f_p$ verifies that $p \in
\mathrm{int(CH}(f_l,f_p,f_r))$.

\begin{prop}
Given a set $F$ of $n$ light sources with limited illumination
range and a point $p$ in the plane, the algorithm MER-Point
computes the MER of $p$ and a $\mathrm{CET}(p)$ in
$\mathcal{O}(n)$ time.
\end{prop}

\emph{Proof}: Let $F$ be a set of $n$ light sources. The distances
from $p$ to all the light sources can be computed in linear time.
Computing the median also takes li\-ne\-ar time \cite{MB73}, as
well as splitting $F$ in two halves. Checking if $p \in
\mathrm{int} (\mathrm{CH}(F')), F'\subseteq F$, is linear on the
number of light sources in $F'$. So the total time for this
logarithmic search is $\mathcal{O}(n + \frac{n}{2} + \frac{n}{4}+
\frac{n}{8}+...) = \mathcal{O}(n)$. Therefore, we find the closest
embracing site for $p$ in linear time. So this algorithm computes
the MER of $p$ and a $\mathrm{CET}(p)$ in total $\mathcal{O}(n)$
time. \hfill $\square$

The decision problem is trivial after the MER of $p$ is computed.
Point $p$ is \mbox{1-well} illuminated if the given illumination
range is greater or equal to the MER of $p$.

\section{Minimum Embracing Range of a Line Segment}

In this section we compute the MER, $r_m$, of a line segment, this
is, we compute the minimum illumination range needed to
\mbox{1-well} illuminate a line segment with a set of light
sources. Without loss of generality, suppose that a line segment
$s = \overline{p_lp_r}$ is an horizontal line segment and that
$p_l$ and $p_r$ are respectively the leftmost and the rightmost
points of $s$. 
Since our solution uses the Parametric Search technique due to
Megiddo \cite{NM79,NM83}, we first concentrate on the following
decision problem: given a range $r>0$, is $s$ \mbox{1-well}
illuminated by a set $F$ of $n$ light sources?

The algorithm in this section decides if a line segment $s$ is
\mbox{1-well} illuminated knowing that the light sources of $F$ have
illumination range $r$. The idea behind this algorithm is to split $s$
into several open segments and check if $r$ is enough to \mbox{1-well}
illuminate all of them. We will show that, if two consecutive open
segments are \mbox{1-well} illuminated, the point in between also
is. Hence, if all segments and both extreme points are \mbox{1-well}
illuminated, so is the segment $s$.

Let us first introduce some notation. We know that each light source
$f_i \in F$ illuminates a circle centered at itself with radius $r$,
$C(f_i,r)$, and that each circle can intersect $s$ in at most two
points. Since we have $n$ light sources, there are $n$ such circles and
at most $2n$ intersection points between $s$ and the circles.  Let $I$
be the set containing the points $p_l=i_0$ and $p_r=i_m$ and all the
sorted intersection points according to their $x$-coordinate $i_1\dots
i_{m-1}$.
If $i_k$ is an intersection point between $C(f_i,r)$ and $s$, let
$f(i_k)$ be $f_i$.  If $i_k \in I$ is to the left (resp. right) of
$f(i_k)$, it is called the leftmost (rightmost) intersection
point. Let also $s_k = (i_{k-1},i_k)$, with $i_{k-1}, i_k \in I$
be the open segment between the intersection points $i_{k-1}$ and
$i_k$, for $k=1, \ldots, m$. Note that the light sources
illuminating $s_k$ are the same for all points in $s_k$ since its
endpoints are consecutive intersections of $I$.  The function
$\mathcal{F}(s_k)$ returns the set of the light sources that
illuminate $s_k \subseteq s$ with range $r$. Knowing that
$\mathcal{F}(s_1) = \mathcal{F}(i_0)$, the function is recursively
defined for $k = 1, \ldots, m-1$ as follows.

\begin{displaymath}
\mathcal{F}(s_{k+1}) = \left\{ \begin{array}{ll}
\mathcal{F}(s_k) \cup \{f(i_k) \}, & \textrm{if $i_k$ is a leftmost intersection}\\
\mathcal{F}(s_k) \setminus \{f(i_k) \}, & \textrm{if $i_k$ is a rightmost intersection}\\
\end{array} \right.
\end{displaymath}

In the Figure \ref{MGV3} there is an example of what happens when the
intersection point $i_k \in I$ is the leftmost point or the rightmost
point.

\begin{figure}[!h]
\centering
\includegraphics[scale=0.8]{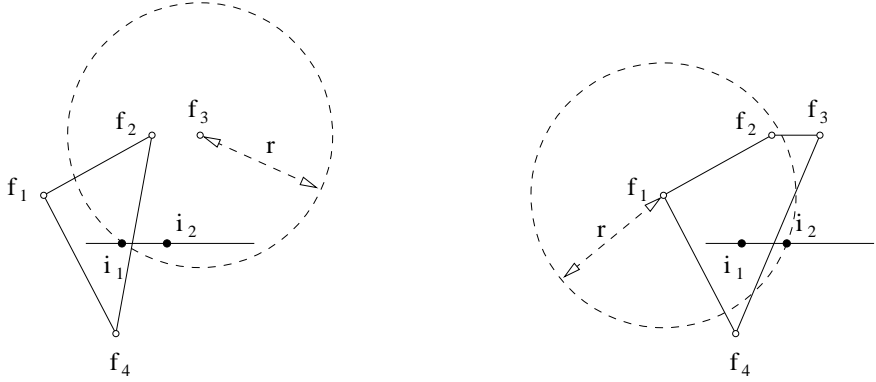}
\caption{(Left) Point $i_1 \in I$ is the leftmost intersection point
between $\overline{p_lp_r}$ and the circle $C(f_3,r)$. The segment
$s_1 = (p_l, i_1)$ is \mbox{1-well} illuminated by the set
$\mathcal{F}(s_1) = \{f_1,f_2,f_4 \}$. (Right) $\mathcal{F}(s_2) =
\mathcal{F}(s_1) \cup \{ f_3\}$. Point $i_2 \in I$ is the
rightmost intersection point between the circle $C(f_1,r)$ and
$\overline{p_lp_r}$. The segment $s_2 = (i_1, i_2)$ is not
\mbox{1-well} illuminated since $i_2 \not \in \mathrm{int}
(\mathrm{CH} (\mathcal{F}(s_2)))$.} \label{MGV3}
\end{figure}

\begin{lem} \label{LemaIlumSeg}
If two consecutive open segments $s_k$ and $s_{k+1}$ are
\mbox{$1$-well} illuminated then $i_k \in I$ is also
\mbox{$1$-well} illuminated.
\end{lem}

\emph{Proof}:

Suppose that $i_k \in I$ is not \mbox{1-well} illuminated, this is, $i_k
\notin (\mathrm{int}(\mathrm{CH}(\mathcal{F}(s_k))) \cup
\mathrm{int}(\mathrm{CH}(\mathcal{F}(s_{k+1}))))$. Since both open
segments are \mbox{1-well} illuminated, $i_k$ must lie on the boundary
of both convex hulls. Since, by definition of $\mathcal{F}(s_k)$, one of
the two convex hulls is contained in the other, and both segments are
inside the biggest one, $i_k$ is also contained. Hence, $i_k \in I$ is
\mbox{1-well} illuminated.

\hfill $\square$

\begin{lem} \label{LemaIlumInt}

The open segment $s_k$ is \mbox{$1$-well}
illuminated if $i_k$ and $i_{k-1}$ are both inside
 $\mathrm{CH}(\mathcal{F}(s_k))$.
\end{lem}

\emph{Proof}: By definition of $\mathcal{F}(s_k)$, the segment $s_k$ is
\mbox{1-well} illuminated if it is interior to the convex hull of this
set. Since this convex hull is obviously convex, the segment $s_k$ is
\mbox{1-well} illuminated when its endpoints are in the convex hull of
$\mathcal{F}(s_k)$. \hfill $\square$

\begin{thm} \label{TeoIluSeg}
If the endpoints of $s$ as well as all $s_k$, with
$k\in\{1,\dots,m\}$,
 are \mbox{$1$-well} illuminated, then $s$ is \mbox{$1$-well}
 illuminated.
\end{thm}

\emph{Proof}: Follows from Lemmas \ref{LemaIlumSeg} and \ref{LemaIlumInt}.
\hfill $\square$

An efficient algorithm to solve the Decision Problem is then the
following: Check if $i_0$ and $i_m$ are \mbox{1-well} illuminated using
the MER-Point algorithm. For all $k\in [1,\dots,m]$ we compute
$CH(\mathcal{F}(s_k))$ by simply adding or deleting a point to
$CH(\mathcal{F}(s_{k-1}))$ and check whether $s_k$ is \mbox{1-well}
illuminated. If all checks are true, then $s$ is \mbox{1-well}
illuminated; otherwise, it is not.

\begin{prop}
Given a real number $r>0$, a set $F$ of $n$ light sources with
limited illumination range $r$ and a line segment $s$, the
algorithm described above decides if $s$ is \mbox{$1$-well}
illuminated by $F$ in $\mathcal{O}(n \log n)$ time.
\end{prop}

\emph{Proof}: Let $F$, be a set of $n$ light sources with a
limited illumination range $r$. Checking if $p_l$ and $p_r$ are
\mbox{$1$-well} illuminated is linear using the MER-Point
algorithm. Sorting the points in $I$ according to their
$x$-coordinate takes $\mathcal{O}(n \log n)$ time. Computing the
set of light sources that illuminate $p_l$ is linear but
constructing its convex hull takes $\mathcal{O}(n \log n)$ time.
Updating dynamically the convex hull every time we need to add or
remove a light source can also be done in $\mathcal{O}(\log n)$
(amortized) time \cite{GB02}, and checking if both $i_{k-1}, i_k$
are in the CH($\mathcal{F}(s_k)$) takes $\mathcal{O}(\log n)$
time. Since we have at most $2n$ intersection points and we spend
$\mathcal{O} (\log n)$ time on each one, this algorithm decides if
$s$ is \mbox{$1$-well} illuminated by $F$ with range $r$ in
$\mathcal{O}(n \log n)$ time. \hfill $\square$

Note that this procedure still works when each light source has a
different range. Instead of having $n$ circles with the same
radius, we have $n$ circles with different radii. After computing
the intersection points between the $n$ circles and the line
segment, the remaining procedure is exactly the same.

\textbf{The Algorithm}

We have an algorithm to decide if, given a range value, a line segment
is \mbox{1-well} illuminated. In order to compute the minimum range
needed to \mbox{1-well} illuminate the segment, we will apply the
Parametric Search technique due to Megiddo \cite{NM79,NM83}.

Let $F$ be a set of $n$ light sources.

Let $g$ be a monotonic function ($g(x) \geq g(y)$ if
$x>y$) with a root. We want to convert our problem in a monotonic
root-finding problem since we have an efficient decision algorithm
to solve it. Now we define the function $g$ as follows:

\begin{displaymath}
g(\theta) = \left\{ \begin{array}{ll}
0, & \textrm{$s$ is \mbox{1-well} illuminated by $F$ with range $r=\frac{1}{\theta}$}\\
1, & \textrm{$s$ is not \mbox{1-well} illuminated by $F$ with range $r=\frac{1}{\theta}$}\\
\end{array} \right.
\end{displaymath}
We know how to compute if $s$ is \mbox{1-well} illuminated using
the decision algorithm just presented. Our goal to find the
greatest root $\theta^* = \max \{ \theta: g(\theta) = 0 \}$ using
the Parametric Search and once we have it, the MER of $s$ is $r_m
= \frac{1}{\theta^*}$.

Since we solve the decision problem in $O(n\log n)$ time, we can
find this root in $O(n^2\log^2 n)$ time. An small improvement in
performance can be achieved using the following parallel decision
algorithm. First, we lexicography sort all the light sources using
$O(n)$ processors which takes $\mathcal{O}(\log n)$ time. We give
each processor one light source so they compute all intersection
points in constant time. Each processor has, at most, two
intersections and has to check if they are inside the convex hull
of the light sources illuminating them. Since the set of light
sources is lexicography sorted, computing the needed convex hull
takes $\mathcal{O}(\log n)$ time with the help of
$\mathcal{O}(\frac{n}{\log n})$ additional processors. Performing
the checks takes $\mathcal{O}(\log n)$ time. With a total number
of $\mathcal{O}(\frac{n^2}{\log n})$ processors, we can decide if
$s$ is \mbox{$1$-well} illuminated in $\mathcal{O}(\log n)$ time.

\begin{prop}
Given a set $F$ of $n$ light sources in the plane and a line
segment $s$, the Parametric Search computes the $\mathrm{MER}$ of
$s$ in $\mathcal{O}(n^2)$ time.
\end{prop}

\emph{Proof}: The sequential decision algorithm takes $S(n) \in
\mathcal{O}(n \log n)$ time while the one running in parallel requires
$T(n) \in \mathcal{O}(\log n)$ time when using $P(n) \in
\mathcal{O}(\frac{n^2}{\log n})$ processors. So the total time to
evaluate the function $g(\theta)$ and finding its greatest root using
the Parametric Search, as well as computing the MER of $s$, is
$\mathcal{O}(S(n)T(n) \log P(n) + T(n)P(n)) \in \mathcal{O}(n \log n
\times \log n \times \log(\frac{n^2}{\log n}) + \log n \times
\frac{n^2}{\log n}) \in \mathcal{O}(n^2)$ time.  \hfill $\square$

\section{The E-Voronoi Diagram Restricted to a Line Segment}

In the previous section we have computed the minimum illumination
range that a set of light sources must have in order to embrace
all the points of a line segment. Now we go further and compute
the closest embracing site for every point of the line segment
which is equivalent to solve the problem of constructing the
E-Voronoi diagram \cite{ACHMP05} restricted to a line segment $s =
\overline{p_lp_r}$. With this structure, it is possible to make a
query in $\mathcal{O}(\log n)$ time to know the minimum
illumination range needed to embrace a point of $s$. As in the
previous section, we will assume that $s$ is an horizontal line
segment and that $p_l$ and $p_r$ are respectively the leftmost and
rightmost points of $s$. The next definition can be found under
the previous name of \emph{MIR-Voronoi region (MIR-VR)}
\cite{ACHMP05}.

\begin{defi}
Let $F$ be a set of $n$ light sources in the plane. For every
light source $f \in F$, the E-Voronoi region of $f$ with
respect to the set $F$ is the set
$$\mathrm{E}\textrm{-}\mathrm{VR}(f,F)= \{x \in \mathbb{R}^2: f \in F \textrm{ is the
closest embracing site for } x \}.$$
\end{defi}

The set of all the E-Voronoi regions is called the E-Voronoi
diagram of $F$ (formerly known as the MIR Voronoi Diagram
\cite{ACHMP05}). An algorithm to compute the E-Voronoi diagram of $F$
restricted to a segment $s$ follows.

For each light source $f \in F$, we perform a sweep
searching for the points of $s$ that belong to the $\mathrm{E}
\textrm{-} \mathrm{VR} (f,F)$. When the sweeping for $f$ is done, we
have computed all the components of the $\mathrm{E} \textrm{-}
\mathrm{VR} (f,F)$ restricted to $s$ (note that the E-Voronoi region of
a light source is not always connected \cite{ACHMP05}). When the
sweeping is done for all the light sources of $F$, we have computed the
E-Voronoi diagram of $F$ restricted to $s$.

Let us start computing the E-Voronoi region of a light source $f
\in F$ restricted to $s$. Let $p_f$ be the closest point on $s$ to
the source $f$. We will sweep from left to right (from $p_f$ to
$p_r$) and then from right to left (from $p_f$ to $p_l$). When we
are moving along $s$, we change from one E-Voronoi region to
another when the point has two closest embracing sites or the
point reaches the border of the convex hull of its current closest
embracing set. In the first case, this corresponds to the
intersection between $s$ and the perpendicular bisectors between
the two closest embracing sites of the point. For that reason, we
compute the intersection points between $s$ and the perpendicular
bisectors between the light sources in $F \backslash \{ f\}$ and
$f$. We keep these intersection points as well as $p_l, p_f$ and
$p_r$ sorted by the $x$-coordinate in two lists (one for each type
of sweeping), $L_1 = \{p_f, \ldots, p_r \}$ and $L_2 = \{p_f,
\ldots, p_l\}$. The intersection points between $s$ and the convex
hull of the closest embracing set of a point will be added to this
list during the sweeping. As the $\mathrm{E} \textrm{-}
\mathrm{VR} (f,F)$ may not be connected, while sweeping $s$ we
might cross several of its components. So to catch up where each
component starts or ends, we say that $p_i \in s$ is a starting
(ending) point of the $\mathrm{E} \textrm{-} \mathrm{VR} (f,F)$ if
it is the point where a component of the $\mathrm{E} \textrm{-}
\mathrm{VR} (f,F)$ starts (ends). We will only explain the
sweeping from left to right using the list $L_1$ (the sweeping
from right to left is a mirror of this one).

For each $p_i \in L_1$ starting on $p_i=p_f$, we compute the
convex hull of all light sources inside the circle with radius
$d(f,p_i)$. We use $H(f,p_i)$ for this hull. Note that $f$ is on
its boundary and call $H^*(f,p_i)$ the obtained convex hull when
deleting $f$ (the two edges adjacent to $f$ are called
\emph{support lines}). We will look for the points $p_i \in s$
such that $p_i \in \mathrm{int} (H(f,p_i))$ but $p_i \notin
\mathrm{int} (H^*(f,p_i))$, this is, the points $p_i\in s$ such
that $f$ is their closest embracing site. The following lemmas
give the clues to the discretization of the sweeping.

\begin{figure}[!h]
\centering
\includegraphics[scale=0.8]{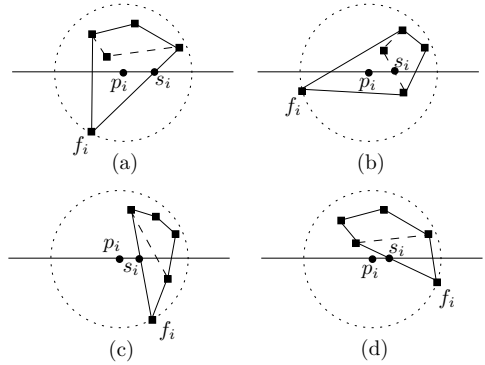}

\caption{Four possible situations for point $p_i$ concerning the
location of $H^*(f,p_i)$ represented by a dashed line and $H(f,p_i)$
represented by a solid line. In cases (a) and (b), the segment between
$p_i$ and $s_i$ is in $\mathrm{E} \textrm{-}\mathrm{VR}(f,F)$. In cases
(c) and (d), $p_i$ is not in the E-Voronoi region of $f$ but $\mathrm{E}
\textrm{-}\mathrm{VR}(f,F)$ starts after the point $s_i$ in both cases.}

\label{MERSeg}
\end{figure}

\begin{lem} \label{LemMER4}
Given a point $p_i \in L_1$ and the light source $f \in F$, let
$p_i \in \mathrm{int}$ $(H^*(f,p_i))$. If  $p_i \in \mathrm{E}
\textrm{-} \mathrm{VR} (f,F)$ then $p_i$ is the ending point of a
component of the $\mathrm{E} \textrm{-} \mathrm{VR} (f,F)$
restricted to $s$.
\end{lem}

\emph{Proof}: Given a point $p_i \in L_1$ and the light source $f
\in F$, let $p_i \in \mathrm{int}(H^*(f,p_i))$. At first, $p_i \in
\mathrm{int}(H^*(f,p_i))$ might suggest that $p_i \notin
\mathrm{E} \textrm{-} \mathrm{VR} (f,F)$ but this may also suggest
that we have just crossed a perpendicular bisector and now there
is another light source in $F$ that also \mbox{1-well} illuminates
$p_i$. So if $p_i \in \mathrm{E} \textrm{-} \mathrm{VR} (f,F)$
then $p_i$ has two closest embracing sites. This means that the
last points on the left of $p_i$ are also in the same component of
the $\mathrm{E} \textrm{-} \mathrm{VR} (f,F)$ as $p_i$, but the
points on its right are in a component of another E-Voronoi
region. So $p_i$ is the ending point of a component of the
$\mathrm{E} \textrm{-} \mathrm{VR} (f,F)$ restricted to $s$.
\hfill $\square$

\begin{lem} \label{LemMER1}
Given two consecutive points $p_i, p_{i+1} \in L_1$ and the light
source $f \in F$, let $p_i \in \mathrm{int} (H(f,p_i))$ but
$p_i \notin \mathrm{int} (H^*(f,p_i))$. If one of the support
lines intersects $s$ between $p_i$ and $p_{i+1}$ on $s_i \in s$,
then $[p_i,s_i)$ is a segment contained in the $\mathrm{E}
\textrm{-}\mathrm{VR}(f,F)$ restricted to $s$.
\end{lem}

\emph{Proof}: Given two consecutive points $p_i, p_{i+1} \in L_1$
and the light source $f \in F$, if $p_i \in \mathrm{int}
(H(f,p_i))$ and $p_i \notin \mathrm{int} (H^*(f,p_i))$ then it
is clear that the point is \mbox{1-well} illuminated by $f$ so
$p_i \in \mathrm{E} \textrm{-}\mathrm{VR}(f,F)$. We know that
between $p_i$ and $p_{i+1}$ we do not cross any perpendicular
bisector between $f$ and the other light sources in $F$ so if
the $\mathrm{E} \textrm{-}\mathrm{VR}(f,F)$ ends before
$p_{i+1}$, that is because the set $H(f,p_i)$ is not an
embracing set for all the points in the segment $\overline{p_i
p_{i+1}}$. Let $s_i \in s$ be the intersection point between $s$
and one of the support lines, so the segment between $p_i$ and
$s_i$ is \mbox{1-well} illuminated by the set $H(f,p_i)$ (see
Figure \ref{MERSeg}(a)). On the other hand, $\overline{s_ip_{i+1}}
\notin H(f,p_i)$ so it is not \mbox{1-well} illuminated by
$f$. This means that $[p_i,s_i)$ is a segment contained in the
E-Voronoi region of $f$ restricted to $s$. \hfill $\square$

\begin{lem} \label{LemMER2}
Given two consecutive points $p_i, p_{i+1} \in L_1$ and the light
source $f \in F$, let $p_i \in \mathrm{int} (H(f,p_i))$ but
$p_i \notin \mathrm{int} (H^*(f,p_i))$. If $H^*(f,p_i)$
intersects $s$ between $p_i$ and $p_{i+1}$, let $s_i \in s$ be the
leftmost intersection point. Then $[p_i,s_i)$ is a segment
contained in the $\mathrm{E} \textrm{-}\mathrm{VR}(f,F)$
restricted to $s$. Otherwise if $p_i$ is the leftmost intersection
point between $H^*(f,p_i)$ and $s$ then $p_i$ is the ending
point of a component of the $\mathrm{E} \textrm{-} \mathrm{VR}
(f,F)$ restricted to $s$.
\end{lem}

\emph{Proof}: Given two consecutive points $p_i, p_{i+1} \in L_1$
and the light source $f \in F$, if $p_i \in \mathrm{int}
(H(f,p_i))$ and $p_i \notin \mathrm{int} (H^*(f,p_i))$ then
$p_i \in \mathrm{E} \textrm{-}\mathrm{VR}(f,F)$. If
$H^*(f,p_i)$ intersects $s$ between $p_i$ and $p_{i+1}$ then
$H(f,p_i)$ is not an embracing set for the whole segment
$\overline{p_ip_{i+1}}$ (see Figure \ref{MERSeg}(b)). Let $s_i \in
s$ be the leftmost intersection point between $s$ and the set
$H^*(f,p_i)$, only $[p_i,s_i)$ is a segment contained in the
E-Voronoi region of $f$ restricted to $s$ because
$\overline{s_ip_{i+1}} \in H^*(f,p_i)$, so it is not
\mbox{1-well} illuminated by $f$. If $p_i$ is the leftmost
intersection between $H^*(f,p_i)$ and $s$ then $p_i$ is already
the ending point of a component of the $\mathrm{E} \textrm{-}
\mathrm{VR} (f,F)$ restricted to $s$. \hfill $\square$

\begin{lem} \label{LemMER3}
Given two consecutive points $p_i, p_{i+1} \in L_1$ and the light
source $f \in F$, let $p_i \notin \mathrm{int}(H(f,p_i))$ and
$p_i \notin \mathrm{int}(H^*(f,p_i))$. If one of the support
lines intersects $s$ between $p_i$ and $p_{i+1}$, let $s_i \in s$
be the leftmost intersection point. Then $s_i$ is the starting
point of a component of the $\mathrm{E} \textrm{-} \mathrm{VR}
(f,F)$ restricted to $s$. Otherwise if $p_i$ is the rightmost
intersection point between the support lines and $s$ then $p_i$ is
the ending point of a component of the $\mathrm{E} \textrm{-}
\mathrm{VR} (f,F)$ restricted to $s$.
\end{lem}

\emph{Proof}: Given two consecutive points $p_i, p_{i+1} \in L_1$
and the light source $f \in F$, if $p_i \notin
\mathrm{int}(H(f,p_i))$ and $p_i \notin
\mathrm{int}(H^*(f,p_i))$ then we know that $p_i \notin
\mathrm{E} \textrm{-} \mathrm{VR}(f,F)$. Nevertheless, if one of
the support lines intersects $s$ between $p_i$ and $p_{i+1}$ then
there are points of $s$ between $p_i$ and $p_{i+1}$ whose
embracing set is $H(f,p_i)$, this is, these points belong to the
E-Voronoi region of $f$. Let $s_i \in s$ be the leftmost
intersection point between $s$ and the support lines (see Figures
\ref{MERSeg}(c) and \ref{MERSeg}(d)). Point $s_i$ is on the border
of $H(f,p_i)$ which means that there will be interior points to
the $H(f,p_i)$ on $s$ after $s_i$. These points are
\mbox{1-well} illuminated by $f$ and $H(f,p_i)$ is their
embracing set so they belong to the E-Voronoi region of $f$.
Thus the $\mathrm{E} \textrm{-} \mathrm{VR}(f,F)$ restricted to
$s$ begins after the point $s_i \in s$. If $p_i$ is the rightmost
intersection point between the support lines and $s$ then
$\overline{p_ip_{i+1}} \notin \mathrm{int}(H(f,p_i))$ which
means that $p_i$ is the ending point of a component of the
$\mathrm{E} \textrm{-} \mathrm{VR} (f,F)$ restricted to $s$.
\hfill $\square$

With these four lemmas, we are now able to sweep $s$ and know what
to do every time we stop on $p_i \in L_1$. If $p_i$ is under the
conditions of Lemma \ref{LemMER1} or Lemma \ref{LemMER2} and $p_i
\neq s_i$, then $[p_i,s_i)$ is a segment contained in the
E-Voronoi region of $f$ restricted to $s$. Point $p_i$ is probably
a starting point of a component of the $\mathrm{E} \textrm{-}
\mathrm{VR} (f,F)$ restricted to $s$ and we need to know if it is
so we know the total extension of this component when we reach its
ending point. The sweeping moves on to $s_i \in s$. If $p_i$ is
under the conditions of Lemma \ref{LemMER3} then the sweeping also
moves on to $s_i \in s$. In the case that $p_i$ is itself an
intersection point (as in the case of the lemmas \ref{LemMER4},
\ref{LemMER2} and \ref{LemMER3}), the sweeping moves on to
$p_{i+1} \in L_1$. If $p_i = p_r$ then we have reached the end of
the list $L_1$ and we move to another sweep. It is clear that if
$p_r \in \mathrm{E} \textrm{-} \mathrm{VR} (f,F)$ then $p_r$ is an
ending point of a component of the $\mathrm{E} \textrm{-}
\mathrm{VR} (f,F)$ restricted to $s$. When the sweeping is done
for both lists, we repeat this procedure for another light source
until we have them all studied.

\begin{thm} \label{TeoMER}
Let $s$ be a line segment and $F$ a set of $n$ light sources. The
algorithm just described computes the E-Voronoi diagram of $F$
restricted to $s$ in $\mathcal{O}(n^2 \log n)$ time.
\end{thm}

\emph{Proof}: Let $F$ be a set of $n$ light sources and $s$ a line
segment. For each light source $f \in F$ we have to sweep $s$
stoping at the $\mathcal{O}(n)$ intersection points between $s$
and the perpendicular bisectors between $f$ and all the light
sources in $F \backslash \{ f\}$, as well as another
intersection points computed during the sweeping. These
intersection points have to be sorted which takes $\mathcal{O}(n
\log n)$ time. Then we have to compute two convex hulls and this
can also be done in $\mathcal{O}(n \log n)$ time. For each
intersection point, we have to search for the cases that interest
us in the lemmas \ref{LemMER1}, \ref{LemMER2}, \ref{LemMER3} and
\ref{LemMER4}, as well as update both convex hulls. Searching the
intersections between $s$ and a convex hull or checking if a point
is interior to a convex hull takes $\mathcal{O}(\log n)$ time.
This is also the (amortized) time spent on dynamically updating
the convex hull \cite{GB02}. So the sweeping for each light source
takes $\mathcal{O}(n \log n)$ time and it computes the restricted
E-Voronoi region of the light source we are studying at the
moment. Since we have $n$ light sources, the algorithm computes
the E-Voronoi diagram of $F$ restricted to $s$ in $\mathcal{O}(n^2
\log n)$ time. \hfill $\square$

Using this algorithm, we compute E-Voronoi diagram of $F$
restricted to $s$ and we can also compute the MER of $s$ using the
following proposition. This technique takes $\mathcal{O}(n^2 \log
n)$ time but it is simpler than the one presented in the previous
section.

\begin{prop}
Let $s$ be a line segment and $F$ a set of $n$ light sources. The
$\mathrm{MER}$ of $s$ is given by the biggest distance between a
light source $f \in F$ and one of the extremes of its E-Voronoi
region restricted to $s$.
\end{prop}

\emph{Proof}: Given a line segment $s$, the MER of $s$ is the
biggest distance between a point of $s$ and its closest embracing
site. By definition, a point is in the E-Voronoi region of $f
\in F$ if $f$ is its closest embracing site. If we compute the
biggest distance between a light source $f$ and a point of $s$
in its E-Voronoi region, we get the minimum illumination range
needed to \mbox{1-well} illuminate all the points in the
$\mathrm{E} \textrm{-} \mathrm{VR} (f,F)$ restricted to $s$. Let
$t$ be the intersection between $s$ and the $\mathrm{E} \textrm{-}
\mathrm{VR} (f,F)$ and $p_i$ be an interior point of $t$. Assume
that $p_f \in t$ (see Figure \ref{lemaMerDois}(a)), since $p_f$ is
the closest point of $t$ to $f$, the distance between $p_i$ and
$f$ increases when we are moving away from $p_f$ along $t$. So
the minimum illumination range needed to \mbox{1-well} illuminate
$t$ is the biggest distance between one of its extreme points and
$f$. Now suppose that $p_f \notin t$ (see Figure
\ref{lemaMerDois}(b)), this means that the distance $d(f,p_i)$
increases if we move $p_i$ towards one of the extremes of $t$ and
decreases towards the other end. Again, the biggest distance
between $f$ and a point of $t$ is the distance between one of
its extremes and $f$. As each light source $f \in F$ has a
minimum illumination range that can be computed by the distance
between $f$ and one of the extremes of its E-Voronoi region
restricted to $s$, the biggest of these minimum illumination
ranges is the MER of $s$. \hfill $\square$

\begin{figure}[!h]
\centering
\includegraphics[scale=1]{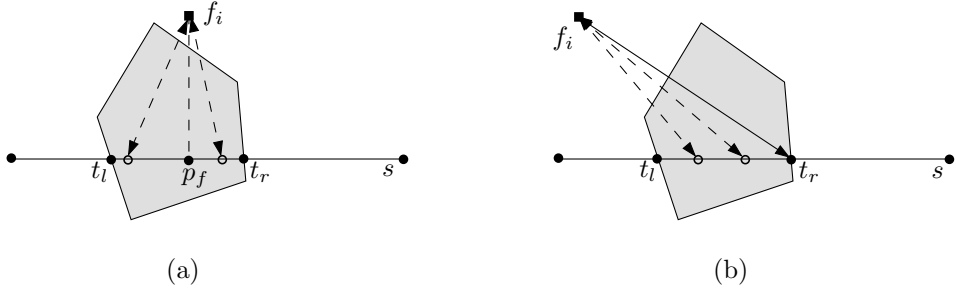}
\caption{(a) The E-Voronoi region of $f \in F$ is the gray area,
$d(f,t_l)$ is the biggest distance between $f$ and a point of
$t$, $d(f, p_i) \leq d(f,t_l), p_i \in \overline{t_lt_r}$. (b)
For a point $p_i \in \overline{t_lt_r}$, $d(f, p_i) \leq
d(f,t_r)$ and the latter is the minimum illumination range
needed to \mbox{1-well} illuminate $t$.} \label{lemaMerDois}
\end{figure}

The most interesting consequence of this algorithm is the
following.

\begin{thm} \label{TeoMER2}
Let $s$ be a line segment and $F$ a set of $n$ light sources. With
a preprocess that can be done in $\mathcal{O}(n^2 \log n)$ time,
one can obtain the MER of a query point $q \in s$ in
$\mathcal{O}(\log n)$ time.
\end{thm}

\emph{Proof}: Let $F$ be a set of $n$ light sources and $s$ a line
segment. We can compute the E-Voronoi diagram of $F$ restricted to
$s$ in $\mathcal{O}(n^2 \log n)$ time using the algorithm
described above. This preprocess allows us to have a structure
that localizes a point $q \in s$ in $\mathcal{O}(\log n)$ time.
Once the point is located, we know what is the E-Voronoi region
that it belongs to, this is, we know what is the closest embracing
site for $q$. The MER of $q$ is given by the distance between $q$
and its closest embracing site. \hfill $\square$

This algorithm and the one in the previous section are also useful
when we want to compute the MER to \mbox{1-well} illuminate a
trajectory or a polygonal line. We can decompose the trajectory in
several line segments and apply one of the algorithms to each
part. This way, we compute a range for each piece and the greatest
range of them all is the MER of the whole polygonal line or
trajectory. An algorithm that shows how to compute all the
different Closest Embracing Triangles for a line segment and their
ranges can be found in \cite{ACH05}.

\section{Conclusions}

The visibility problems solved in this paper consider $n$ light
sources. We presented the linear algorithm \emph{MER-Point} for
computing a CET($p$) and its MER. This algorithm can also be used
to decide if a point in the plane is \mbox{1-well} illuminated. We
also presented a quadratic algorithm to compute the MER of a line
segment in the plane using the Parametric Search. Concerning the
main subject in this paper, we presented another algorithm that
computes the E-Voronoi diagram restricted to a line segment, as
well as its MER. Both algorithms can also be extended to compute
the MER of either open or closed polygonal lines.

} 
\end{document}